\documentstyle[preprint,aps]{revtex}
\tighten
\begin{document}
\draft
\preprint{LPTENS 94/13}
\title{ Solution of the two impurity, two channel Kondo Model}
\author{Antoine Georges and Anirvan M. Sengupta}
\address{
CNRS-Laboratoire de Physique Th\'{e}orique de l'Ecole
Normale Sup\'{e}rieure$^1$
\\
e-mail: georges@physique.ens.fr, sengupta@physique.ens.fr
\\
24, rue Lhomond; 75231 Paris Cedex 05; France}

\date{May,20 1994}
\maketitle
\begin{abstract}
We solve the two-impurity two-channel Kondo model
using a combination of conformal invariance and bosonisation techniques.
The odd-even symmetric case is analysed in detail. The RKKY interaction turns
out to be exactly marginal, resulting in a line of non-Fermi liquid
fixed points. Explicit formulae are given for the critical exponents and for
the finite-size spectrum, which depend continuously on a single parameter.
The marginal line spans a range of values of the RKKY coupling $I$ which
goes from the infinitely strong ferromagnetic point $I=-\infty$
(associated with a 4-channel spin-1 Kondo model) to a finite antiferromagnetic
critical value $I_c>0$ beyond which a Fermi liquid is recovered.
We also find that, when the odd-even symmetry is broken, the marginal line is
unstable for ferromagnetic $I$, while for antiferromagnetic $I$
it extends into a manifold of fixed points.

\vspace{3cm}

\begin{description}
 \item[$^1$]  Unit\'{e} propre du CNRS (UP 701) associ\'{e}e \`{a} l'ENS  et
\`{a} l'Universit\'{e} Paris-Sud
\end{description}

\end{abstract}

\narrowtext
The effect of inter-impurity interactions on quantum single-impurity models
possessing a non-Fermi liquid ground-state is
of crucial importance for the possible experimental realizations of such
systems \cite{exp}. It is also of considerable theoretical interest for
understanding
non-Fermi liquid behaviour in lattice models of correlated fermions
starting from a local point of view \cite{CV,GK}.

The two-impurity Kondo model with two channels of conduction electrons
is one of the simplest model where this problem can be addressed.
For a single impurity, this model is controlled by a non-trivial
fixed point \cite{NB}, resulting in a specific heat coefficient
$C/T$ and susceptibility
$\chi_{imp}$
diverging logarithmically as $T\rightarrow 0$, and a universal finite-size
spectrum
of excitation energies differing from the free-fermion form \cite{AL}.
The corresponding two-impurity model is the simplest situation which brings
in the
competition between the formation of this non-trivial Kondo state and the
ordering of the impurities via the RKKY interaction. It has been
recently studied by numerical renormalization-group methods (NRG)
\cite{IJW,KI1,KI2}.

In this letter, we present an analytic solution of the
low-energy universal properties of this model using a combination of
conformal field theory \cite{AL,AL2} and bosonisation methods
\cite{EK,EKetc}.
We find that the RKKY interaction (as well as other interimpurity
couplings) is a marginal perturbation, giving rise to a continuous family
of non-Fermi liquid fixed points. The finite-size spectrum and the
critical properties vary continuously with the strength of the interaction.
We obtain analytic formulae for this dependence. These results are in
excellent agreement with recent NRG results \cite{IJW,KI1,KI2}. They should be
contrasted with the single-channel case in which, in presence of
particle-hole symmetry, Kondo screening
always dominate over RKKY ordering or vice-versa, resulting in two stable
Fermi liquid fixed points separated by an unstable non-trivial critical
point \cite{JVW,AL2}.

We formulate the model in terms of left-moving (chiral) fermions
$\psi_{l i\alpha} (x)$ on the
full axis
$-\infty<x<+\infty$. $l=1,2$ is an index labelling the two impurity sites,
$i=1,2$ is a channel index and $\alpha$ a spin index. The hamiltonian is
written as:
\begin{eqnarray}
&H = i v_F \sum_{l i\alpha} \int_{-\infty}^{+\infty} dx
\psi_{l i\alpha}^{\dagger}(x){{\partial}\over{\partial x}}
\psi_{l i\alpha}(x)\\
&+ J_{+} (\vec{S}_1+\vec{S}_2).(\vec{{\cal J}}_1(0)+\vec{{\cal J}}_2(0))\\
&+ J_{m} (\vec{S}_1-\vec{S}_2).(\vec{{\cal J}}_1(0)-\vec{{\cal J}}_2(0))\\
&+ J_{-} (\vec{S}_1+\vec{S}_2).\sum_{i,\alpha\beta}
(\psi_{1 i\alpha}^{\dagger}(0) {{\vec{\sigma}_{\alpha\beta}}\over{2}}
\psi_{2 i\beta}(0)
+\psi_{2 i\alpha}^{\dagger}(0) {{\vec{\sigma}_{\alpha\beta}}\over{2}}
\psi_{1 i\beta}(0))\\
&+ I \vec{S}_1.\vec{S}_2
\end{eqnarray}
In these formulas,
$\vec{{\cal J}}_l(x)=\sum_{i,\alpha\beta}
\psi_{l i\alpha}^{\dagger}(x) {{\vec{\sigma}_{\alpha\beta}}\over{2}}
\psi_{l i\beta}(x)$
denotes the spin-current at position $x$ for species $l$. Our notations follow
closely those of Ref.\cite{AL2}. Alternatively, one could work in the
even/odd basis $\psi_{e,o}$ with respect to the midpoint between impurities.
The combinations $\psi_{1,2}$ correspond to $(\psi_e\pm\psi_o)/\sqrt{2}$,
respectively, and a parity transformation amounts to exchange the indices
$l=1,2$ for both impurity spins and conduction electrons. In order to make
contact with the couplings $\Gamma_{e,o,m}$ used in Ref.\cite{IJW}, let us note
the identifications:
$J_m\propto\Gamma_m$, $J_{+}\propto \Gamma_e+\Gamma_o$,
$J_{-}\propto \Gamma_e-\Gamma_o$.

We shall start by identifying the global symmetries of the hamiltonian.
For most of this paper we shall concentrate on the case $J_{-}=0$,
corresponding to a hamiltonian invariant under odd-even exchange
($\Gamma_e=\Gamma_o$).
$H$ has a higher symmetry in that case, with independent charge and
channel (or `flavour') transformations allowed for $l=1,2$:
\begin{eqnarray}
&\psi_{l i\alpha}\rightarrow e^{i\theta_l} \psi_{l i\alpha}\\
&\psi_{l i\alpha}\rightarrow \sum_j U^{(l)}_{ij} \psi_{l j\alpha}\,\,,\,\,
U^{(l)}\in SU(2)
\label{flavsym}
\end{eqnarray}
When $J_{-}\neq 0$, only $\theta_1=\theta_2$ and $U^{(1)}=U^{(2)}$ are
allowed.
The symmetry of the spin sector is conveniently discussed by looking
first at the case of two {\it decoupled} impurities (each one interacting
with two conduction channels) obtained by
setting $I=0$ and $J_{+}=J_m$ in addition to $J_{-}=0$. In that case, $H$
has independent spin-rotation symmetry:
\begin{equation}
\psi_{l i\alpha}\rightarrow \sum_{\beta} V^{(l)}_{\alpha\beta}
\psi_{l i\beta}\,\,,\,\,S_l^{a}\rightarrow \sum_{b}
R^{ab}(V^{(l)}) S_l^{b}
\label{spinsym}
\end{equation}
where $R^{ab}(V)=1/2 tr(\sigma^{a}V\sigma^{b}V^{\dagger})\,\,,\,\,(a,b=x,y,z)$
is the adjoint representation of $V\in SU(2)$. Hence, two decoupled
2-channel Kondo models have global symmetry
$(SU(2)_{spin}\otimes SU(2)_{flavour}\otimes U(1)_{charge})^2$.
Coupling the two impurities ($I\neq 0, J_{+}\neq J_m$) while keeping
$J_{-}=0$ leaves unchanged the independent charge and flavour symmetries,
but reduces the spin symmetry to the diagonal $SU(2)$ corresponding to
$V^{(1)}=V^{(2)}$ in Eq.(\ref{spinsym}).

At a fixed point, these global symmetries are promoted to local conformal
symmetries \cite{PG}.
For decoupled impurities, the symmetry algebra consists in two
copies of a product of
Kac-Moody algebra for spin, channel and charge:
$(\widehat{SU}_2(2)_{s}\otimes \widehat{SU}_2(2)_{f}\otimes
\widehat{U}(1)_{c})^2$.
($\widehat{SU}_k(2)$ stands for the level-$k$
$SU(2)$ Kac-Moody algebra, corresponding to the commutation relations of
the sum of $k$ independent $SU(2)$ currents).
When coupling the impurities with $J_{-}=0$, the
diagonal $SU(2)$ symmetry of the spin sector gives rise to a
$\widehat{SU}_4(2)$ algebra. The generators of this algebra are the
sum of the generators of the two $\widehat{SU}_2(2)_s$ for each impurity,
that is the sum of
the spin currents $\vec{{\cal J}}_1(x)+\vec{{\cal J}}_2(x)$.
Hence, we must understand how the
product $\widehat{SU}_2(2)_{s}\otimes \widehat{SU}_2(2)_{s}$ can be
decomposed into $\widehat{SU}_4(2)_{s}$ plus some residual degrees of
freedom. The answer is given by the so-called coset construction \cite{GKO}:
\begin{equation}
\widehat{SU}_2(2)_{s}\otimes \widehat{SU}_2(2)_{s}
=\widehat{SU}_4(2)_{s} \otimes A(2,2)
\label{coset}
\end{equation}
The algebra $A(2,2)$ turns out to be a $N=1$ superconformal unitary model
\cite{PG,A22}
corresponding to the $m=4$ member of the discrete series with
central charge $c={{3}\over{2}}(1-{{8}\over{m(m+2)}})$, and thus has $c=1$.
This construction generalizes to the two-channel case the one made by
Affleck and Ludwig in their solution of the one-channel two impurity
problem \cite{AL2}. There, the coset construction is
$\widehat{SU}_1(2)_{s}\otimes \widehat{SU}_1(2)_{s}
=\widehat{SU}_2(2)_{s} \otimes A(1,1)$, where the algebra $A(1,1)$ is
actually an Ising model with $c=1/2$.

This coset construction can be understood more explicitly when dealing
with spin currents (i.e for the adjoint representation of the algebra).
Let us first recall \cite{PG} that the $\widehat{SU}_2(2)$ spin current
${\cal J}_l^a(x)$ ($a=x,y,z$) for a given $l=1,2$ can be represented
in terms of
three Majorana (i.e real) fermions $\chi_l^{x,y,z}$ as follows:
\begin{equation}
{\cal J}_l^a(x)=i \epsilon_{abc} \chi_l^b \chi_l^c
\end{equation}
This is particularly transparent when using the Emery-Kivelson
bosonisation approach
to the two-channel Kondo model \cite{EK,EKetc}, in which case
$\chi^x=sin\Phi_s, \chi^y=cos\Phi_s, \chi^z=cos\Phi_{sf}$ where $\Phi_s,
\Phi_{sf}$ are the boson field introduced in ref.\cite{EK} corresponding
to spin and `spin-flavour'
degrees of freedom.
Here, we are dealing with two copies of $\widehat{SU}_2(2)$ and hence with
six Majorana fermions. We combine them into three Dirac fermions and bosonize
these new degrees of freedom as:
\begin{equation}
\chi_1^a(x) + i \chi_2^a(x) =
e^{i\Phi_{a}(x)}\,\,,\,\,a=x,y,z
\label{Dirac}
\end{equation}
Our notations
are such that the free boson correlator reads:
$<\Phi(r)\Phi(0)>=-ln(r)$, so that $e^{ik\Phi}$ has dimension
$k^2/2$. In terms of these fields, the
total spin current
corresponding to the diagonal $\widehat{SU}_4(2)$ algebra
reads:
${\cal J}^x={\cal J}_1^x+{\cal J}_2^x=cos(\Phi_y-\Phi_z)$ (and cyclic
permutations). It is convenient to introduce three linear combinations
of boson fields as follows:
\begin{equation}
\Phi = {{1}\over{\sqrt{3}}} (\Phi_x+\Phi_y+\Phi_z)\,\,,\,\,
\mu = {{1}\over{\sqrt{2}}} (\Phi_x-\Phi_y)\,\,,\,\,
\nu = {{1}\over{\sqrt{6}}} (\Phi_x+\Phi_y-2\Phi_z)
\label{bosons}
\end{equation}
In term of these combinations, the components of the total spin current
read:
\begin{equation}
{\cal J}^x=cos({{\mu-\sqrt{3}\nu}\over{\sqrt{2}}})\,\,,\,\,
{\cal J}^y=cos({{\mu+\sqrt{3}\nu}\over{\sqrt{2}}})\,\,,\,\,
{\cal J}^z=cos(\sqrt{2}\mu)
\end{equation}
Note that $\Phi$ does not enter these expressions. Hence the two bosons
$\mu,\nu$ are sufficient to describe the
$\widehat{SU}_4(2)$ algebra (as expected from its central charge $c=2$) and
$\Phi$ corresponds to the residual $A(2,2)$ degree of freedom ($c=1$).
(The central charge $c=3/2+3/2$ has thus been distributed as $c=2+1$ in
the coset construction Eq.(\ref{coset}).

Thus, a very useful explicit realization of the algebra $A(2,2)$
as a free field
theory of a single compact chiral boson $\Phi$ has been found.
Since $\Phi_{x,y,z}$ have periodicity $2\pi$, the radius of $\Phi$
is found from eq.(\ref{bosons}) to be
$R=\sqrt{3}$, which
means that $\Phi$ and $\Phi+2\pi R$ are identified. However, this construction
based on currents in the adjoint representation does not reveal the
full structure of the $\widehat{SU}_2(2)$ and $A(2,2)$
algebra. Indeed, the three Majorana fermions involved in the currents of
$\widehat{SU}_2(2)$ are not fully independent. For example, if we change
their boundary
conditions from periodic to antiperiodic, all three must be changed
simultaneously so that the boundary condition on the current remain periodic.
The boundary conditions on the two sets of Majorana fermions $\chi_l^a$,
$l=1,2$ are independent however, which means that the fermions constructed
in Eq.(\ref{Dirac}) are not strictly speaking Dirac fermions but rather that
$\Phi$ and $-\Phi$ must be
identified. $A(2,2)$ is thus really a (chiral) `orbifold' theory.
This is of crucial importance in determining the operator content of this
algebra, and thus the finite size spectrum of the present model.
It implies that, in addition to the operators
$e^{\pm i (n\sqrt{3}+m/2\sqrt{3})\Phi}$ and $\partial^n\Phi$ (with
$n,m$ integers), $A(2,2)$ contains two operators of dimension $1/16$ which
do not have a simple boson representation.
These operators are analogous to the (twist) operator associated with the
order parameter for the case of the Ising model, which change the boundary
condition of the Majorana fermion from periodic to antiperiodic \cite{foot1}.
For convenience, the full set of primary operators of the $A(2,2)$
algebra and their boson
representation (when it exists) is given in Table 1.

We now consider the effect of turning on the `RKKY' coupling $I$,
and/or setting $J_{+}\neq J_{m}$, starting
from decoupled impurities (but keeping $J_{-}=0$, i.e $\Gamma_e=\Gamma_o$).
An order by order perturbative calculation in these couplings
(e.g of the free energy)
involves correlation functions of the operator $\vec{S_1}.\vec{S_2}$
which is a product of spin correlators at the decoupled impurities fixed
point.
As will be detailed in a longer publication \cite{AAcoming}, it can be shown
that, for the purpose of calculating these correlations, the impurity
spin $\vec{S_1}$ (resp. $\vec{S_2}$) can be replaced by
$a_1\vec{\chi_1}$ (resp. $a_2\vec{\chi_2}$), where $a_{1,2}$ are local
real fermions needed to ensure proper commutations. Hence, the
perturbing term of lower dimension associated with the RKKY interaction
reads
$\int dt a_1 a_2 \vec{\chi_1}.\vec{\chi_2}$. In the bosonic language above,
this translates into an induced boundary term in the
$A(2,2)$ sector of the hamiltonian:
\begin{equation}
H_{A(2,2)}={v_F \over 4} \int dx (\Pi(x)-{{\partial \Phi}\over{\partial x}})^2
+\tilde{I} (d^{+}d-{{1}\over{2}}) {{\partial \Phi}\over{\partial x}}(0)
\label{hamphi}
\end{equation}
where we have set $d^{+}\equiv (a_1+ia_2)/\sqrt{2}$,
$\Pi$ is the field conjugate to $\Phi$  and $\tilde{I}$ is
some (non-universal) function of $I$ and $J_{+}-J_{m}$.
Hence the RKKY coupling is associated
with a dimension $1$ operator and is
an exactly marginal perturbation.

The hamiltonian (\ref{hamphi}) is very similar to the X-ray
edge hamiltonian (in presence of the charged core) in the bosonised form
\cite{ScSc}.
In the X-ray edge problem, the boundary term gives a phase-shift, which
changes the dimensions of the operators $e^{\pm ik\Phi}$ from ${k^2 \over 2}$
to ${{1}\over{2}}(k\pm {\delta\over\pi})^2 $. This can be obtained by
formally fusing $e^{\pm ik\Phi}$ with
$U_{\delta}\equiv \exp( i{{2\delta}\over{\pi}}(d^{+}d-1/2)\Phi(0))$.
This is the
analogue of the "fusion hypothesis" introduced by Affleck and Ludwig for
the Kondo problem \cite{AL,AL2}.
This fusion rule
can be generalised to the present problem (with $\delta$ some non-universal
function of $I$ and $J_{+}-J_{m}$)
provided we know how to generalise it for the operators of
dimension $({1 \over 16})$. It turns out \cite{AAcoming}
that the fusion gives back the
$({1 \over 16})$ operators plus operators of dimension
${1 \over 16}+{1 \over 2}$ and conformal descendants of these two operators.

Using this method, we have derived
the finite-size spectrum
of the model for $I\neq 0$, $J_{+}\neq J_{m}$ (but $J_{-}=0$)
as a function of the single
parameter $\delta$, starting from the (known \cite{AL}) finite-size spectrum
of two decoupled two channel Kondo models.
This is displayed in Table 2 (where the parameter
$x\equiv 2\sqrt{3}{\delta \over \pi}$ has been used).
Note that the ground-state is the triplet of lowest energy for $x>0$
(ferromagnetic coupling), and
the singlet of lowest energy for $x<0$ (antiferromagnetic coupling).
Accordingly,
the normalised excitation energy of a given state,
${{L\Delta E} \over {\pi v_F}}$
(with $L$ the radial length of the bulk system and $v_F$ the Fermi velocity),
is obtained from the total dimension $\Delta_{tot}$ given
in Table 2 by the formula:
\begin{eqnarray}
{{L\Delta E} \over {\pi v_F}} = \Delta_{tot} - {1 \over 3}
-{(1-x)^2 \over 24} \,\,,\,\, (for\,\, x<0)\\
{{L\Delta E} \over {\pi v_F}} = \Delta_{tot} -
{3\over 8}(1+{x\over 3})^2\,\,,\,\, (for\,\,
x>0)
\label{energy}
\end{eqnarray}
These formulae are in excellent agreement with recent
numerical renormalization group
results of
K.Ingersent and B.Jones obtained
by the numerical renormalization group method \cite{KI2}.
They correspond to a one-parameter family
of (non-Fermi liquid) fixed
points (when symmetry between odd and even channels is preserved) which can be
explored by varying either $I$ or $J_{+}-J_{m}$.
One of the endpoints of this line is the strong ferromagnetic fixed point
of Ref.\cite{IJW}, obtained for $I\rightarrow -\infty$ (corresponding to
$x=1$). At this point, the impurity spins bind into a triplet state, and
the spectrum of the 4-channel
spin-1 Kondo problem is found, as conjectured in Ref.\cite{IJW}.
The $A(2,2)$ operator associated with each state at this fixed-point is
also displayed in Table 2.
The marginal line includes the point corresponding to two
decoupled Kondo problems ($x=0$) and extends over to a region of
antiferromagnetic
coupling. As discussed below, we expect it to end at $x=\sqrt{6}-3\simeq -.55$
on this side.

The above fusion principle can also be applied to find the dimensional change
of any operator $\cal{O}$ for a non-zero $x$ ($\cal{O}$ transforms as
$U_{\delta}\cal{O}U_{\delta}^{+}$), and hence the operator content at a
given fixed point along the line.
Spin correlations can be obtained from the identification
$S_1^a\pm S_2^a \propto d e^{\pm i\Phi^a} + h.c$, which leads to a singular
behaviour of the uniform susceptibility $\chi_{imp}\propto T^{-\theta(x)}$
on the ferromagnetic side $x>0$ with a continuously varying exponent
$\theta(x)= x(2-x)/3$. Similarly, the staggered susceptibility behaves as
$\chi_{st}\simeq T^{-\theta(-x)}$ on the antiferromagnetic side
$x<0$. $\chi_{st}$ (resp. $\chi_{imp}$) is finite for $x>0$ (resp. $x<0$).
The low-temperature behaviour of the specific heat is governed by
the leading irrelevant perturbations compatible with all
symmetries of the model.
At the decoupled impurities point,
one has two such operators, of dimension $3/2$ which read
$(\vec{S_1}\pm\vec{S_2}).(\vec{\cal{J}_1}\pm\vec{\cal{J}_2})$.
In the bosonised form, they involve
$d e^{-i\sqrt{3}\Phi} +h.c$ and
$d^{+} e^{-i\Phi/\sqrt{3}} \sum_a\xi_a{\cal J}_a +h.c$, where
$\xi_a$ is the adjoint operator of the
$\widehat{SU}_4(2)$ algebra. For a non-zero $x$, they are changed into
$d e^{-i(\sqrt{3}+x/\sqrt{3})\Phi} +h.c$ and
$d^{+} e^{-i\Phi (1-x)/\sqrt{3}} \sum_a\xi_a{\cal J}_a +h.c$,
respectively. Hence the leading contribution to $C/T$ is due to the
latter for $x>0$, and to the former for $x<0$. This leads to
$C/T\simeq T^{-\theta(x)}$ and to a universal ($x$-dependent)
Wilson ratio for $x>0$, while a different behaviour
$C/T\simeq T^{-\alpha(x)}$ with
$\alpha(x)=-x(6+x)/3$ is found for $x<0$.
Note that this yields $\alpha=0$ at
$x=x_c=\sqrt{6}-3$, corresponding to the first operator above becoming
marginal. We expect that, for $x<x_c$, the system flows to the `strong-
antiferromagnetic' (Fermi liquid) fixed point where the two
impurities bind into a
singlet state, as found in \cite{IJW}.

Finally, we report on some results on the effect of
a non-zero coupling $J_{-}$ (even-odd asymmetry).
It can be shown that it gives rise to a
{\it relevant} perturbation on the ferromagnetic side ($x>0$), thus
destabilizing the marginal line in favour of a spin-$1$, $2$-channel
Fermi liquid fixed point, as found in \cite{IJW}. For antiferromagnetic
RKKY interactions ($x<0$), the leading operators generated are marginal,
and the line extends into a surface of non-Fermi liquid fixed points with
continuously varying properties, in agreement with recent NRG
findings \cite{KI1}. These will be analysed in detail in
a forthcoming work \cite{AAcoming}, together with the effect of
particle-hole symmetry
breaking.

\acknowledgments
We are extremely grateful to Barbara Jones, and especially to Kevin Ingersent
for a very fruitful correspondance and for the
communication of recent
NRG results \cite{KI2}. This exchange played a crucial role
in the elaboration
of the present work. We also acknowledge discussions with
E.Kiritsis and Th.Giamarchi, and the hospitality of ICTP (Trieste) where
this work was completed. After completion of this work, we received a preprint
by I.Affleck and A.Ludwig where the analysis of the X-ray edge problem
in terms of boundary-condition changing operators is also performed.

\newpage
\centerline{{\bf Table Captions}}
\noindent
{\bf Table 1}

Primary operators of the $A(2,2)$ algebra. $(\Delta)$ labels an
operator of dimension $\Delta$. The second line displays the bosonic
representation of the operator (when it exists). $NS$ and $R$ stand for
the Neveu-Schwarz and Ramond sectors of the algebra.

\noindent
{\bf Table 2}

Finite-size spectrum of low-lying states. $j$ is the total spin quantum
number, $j_1$ (resp. $j_2$) is the $SU(2)_{flavour}$ quantum number for
$l=1$ (resp. $l=2$), and $Q_1$ (resp. $Q_2$) is the charge.
The sixth column displays the $A(2,2)$ operator associated with each
eigenstate at the decoupled impurities fixed point ($I=0$, i.e $x=0$),
whereas the eigth
column displays the corresponding operator at the strong ferromagnetic
fixed point ($I=-\infty$ i.e $x=1$). The degeneracy of each state is displayed
in the last column, while $\Delta_{tot}$ is the total conformal dimension
at arbitrary $x$
entering the formula for the excitation energy given in the text
(eq.(\ref{energy}).

\newpage

\begin{math}
\begin{array}{|c|c|c|c|c|c|c|c|c|c|}\hline
 & & & & & & & & & \\
(0) & ({{1}\over{24}}) & ({1\over16}) & ({1\over 16}) &
({1\over 6}) & ({3\over 8}) & ({1\over 16}+{1\over 2}) &
({9\over 16}) & ({1\over 6}+{1\over 2}) & (1)\\[.1in] \hline
 & & & & & & & & &\\
1 & e^{\pm i {{\Phi}\over{2\sqrt{3}}}} &
 & &
e^{\pm i {{\Phi}\over{\sqrt{3}}}} &
e^{\pm i {\sqrt{3}\Phi\over2}} &
 & &
e^{\pm i {{2\Phi}\over{\sqrt{3}}}} &
\partial\Phi \\[.1in] \hline
NS & R & NS & R & NS & R &  NS & R & NS & NS \\ \hline
\end{array}
\end{math}
\medskip
\centerline{{\bf Table 1}}
\bigskip

\begin{math}
\begin{array}{|c|c|c|c|c|c|c|c|c|}\hline
j & j_1 &j_2 & Q_1 & Q_2  & A(2,2) & \Delta_{tot}& A(2,2) &
Deg. \\
 & & & & & decoupled & & strong ferro. &
\\[.1in]\hline\hline
0 & 0 & 0 & 0 & 0 & ({{3}\over{8}}) & {{3}\over{8}} (1+{{x}\over{3}})^2 &
({1\over 6}+{1\over 2}) & 1
\\[.1in] \hline
1 & 0 & 0 & 0 & 0 & ({{1}\over{24}}) & {{1}\over{3}}+{{(1-x)^2}\over{24}}
& (0) & 3
\\[.1in] \hline
{{1}\over{2}} & {{1}\over{2}} & 0 & \pm 1 & 0 & ({{1}\over{16}}) &
{{1}\over{2}} & ({{1}\over{16}}) & 8
\\[.1in]
{{1}\over{2}} & 0 & {{1}\over{2}} & 0 & \pm 1 & ({{1}\over{16}}) &
{{1}\over{2}} & ({{1}\over{16}}) & 8 \\[.1in] \hline
0 & {{1}\over{2}} & {{1}\over{2}} & \pm 1 & \pm 1 & (0) &
{{5}\over{8}}+{{x^2}\over{24}} & ({{1}\over{24}}) & 8
\\[.1in]
0 & {{1}\over{2}} & {{1}\over{2}} & \pm 1 & \mp 1 & (0) &
{{5}\over{8}}+{{x^2}\over{24}} & ({1\over 24}) & 8
\\[.1in]\hline
0 & 0 & 0 & \pm 2 & 0 & ({{3}\over{8}}) & {{1}\over{2}}+{{3}\over{8}}
 (1-{{x}\over{3}})^2 & ({1\over 6}) & 2
\\[.1in]
0 & 0 & 0 & 0 & \pm 2 & ({{3}\over{8}}) & {{1}\over{2}}+{{3}\over{8}}
 (1-{{x}\over{3}})^2 & ({1\over 6}) & 2
\\[.1in]
0 & 1 & 0 & 0 & 0 & ({{3}\over{8}}) & {{1}\over{2}}+{{3}\over{8}}
 (1-{{x}\over{3}})^2 & ({1\over 6}) & 3
\\[.1in]
0 & 0 & 1 & 0 & 0 & ({{3}\over{8}}) & {{1}\over{2}}+{{3}\over{8}}
 (1-{{x}\over{3}})^2 & ({1\over 6}) & 3
\\[.1in]\hline
1 & 0 & 0 & \pm 2 & 0 & ({{1}\over{24}}) & {{5}\over{6}}+{{(1+x)^2}\over{24}}
& ({1\over 6}) & 6
\\[.1in]
1 & 0 & 0 & 0 & \pm 2 & ({{1}\over{24}}) & {{5}\over{6}}+{{(1+x)^2}\over{24}}
& ({1\over 6}) & 6 \\[.1in]
1 & 1 & 0 & 0 & 0 & ({{1}\over{24}}) & {{5}\over{6}}+{{(1+x)^2}\over{24}}
& ({1\over 6}) & 9 \\[.1in]
1 & 0 & 1 & 0 & 0 & ({{1}\over{24}}) & {{5}\over{6}}+{{(1+x)^2}\over{24}}
& ({1\over 6}) & 9 \\[.1in]\hline
\end{array}
\end{math}
\medskip
\centerline{{\bf Table 2}}


\newpage
\begin{references}
\bibitem{exp} D.L Cox, {\sl Phys.Rev.Lett.} {\bf 59}, 1240 (1987);
B.Andraka and A.M Tsvelik, {\sl Phys.Rev.Lett.} {\bf 67}, 2886 (1991);
A.Zawadosky and K.Vladar, {\sl Solid State Commun.} {\bf 35}, 217 (1980);
A.Muramatsu and F.Guinea, {\sl Phys.Rev.Lett.} {\bf 57}, 2337 (1986);
D.C Ralph and R.A Buhrman, {\sl Phys.Rev.Lett} {\bf 69}, 2118 (1992).

\bibitem{CV} I. Perakis, C.M. Varma and A. E. Ruckenstein,
{\sl Phys. Rev. Lett} {\bf 70}, 3467 (1993);
T. Giamarchi, C.M. Varma,
A.E. Ruckenstein and P. Nozieres, {\sl Phys. Rev. Lett} {\bf 70}, 3967 (1993);
Q. Si and G. Kotliar {\sl Phys. Rev. Lett} {\bf 70}, 3143 (1993).

\bibitem{GK} A.Georges and G.Kotliar, {\sl Phys.Rev.B} {\bf 45}, 6479 (1992).

\bibitem{NB} Ph. Nozieres and A. Blandin,
{\sl J. Phys. (Paris)} {\bf 41}, 193 (1980);
N. Andrei and C. Destri, {\sl Phys. Rev. Lett.} {\bf 52},
364 (1984); A. M. Tsvelick and P. B. Wiegmann, {\sl J. Stat. Phys.}
{\bf 38}, 125 (1985); P. D. Sacramento and P. Schlottmann, {\sl Phys.
Lett. A} {\bf 142}, 245 (1989).

\bibitem{AL} I.Affleck and A.W.W. Ludwig, {\sl Nucl. Phys. B}
{\bf 352}, 849 (1991); I. Affleck and A.W.W. Ludwig, {\sl Nucl. Phys. B}
{\bf 360}, 641 (1991); I.Affleck, A.W.W Ludwig, H.B Pang and D.L Cox
{\sl Phys.Rev.B} {\bf 45}, 7918 (1992).

\bibitem{IJW} K.Ingersent, B.A Jones and J.W Wilkins {\sl Phys. Rev. Lett.}
{\bf 69}, 2594 (1992).

\bibitem{KI1} K.Ingersent and B.Jones, IBM report RJ 9566 (83572), to
appear in the proceedings of the SCES conference (La Jolla, 1993).

\bibitem{KI2} K.Ingersent and B.Jones, {\sl Private communication and
in preparation}.

\bibitem{AL2} I. Affleck and A.W.W. Ludwig {\sl Phys. Rev. Lett.}
{\bf 68}, 1046 (1992).

\bibitem{EK} V. J. Emery and S. Kivelson,
{\sl Phys. Rev. B} {\bf 46}, 10812  (1992).

\bibitem{EKetc} See also A.Sengupta and A.Georges,
{\sl Phys.Rev.B} {\bf 49}, 10020 (1994) ;
D.G Clarke, T.Giamarchi and B.Shraiman
{\sl Phys.Rev.B} {\bf 48}, 7070 (1993).

\bibitem{JVW} B.Jones, C.M.Varma and J.W.Wilkins, {\sl Phys.Rev.Lett.}
{\bf 61}, 125 (1988).

\bibitem{PG} For a general reference on conformal field theory, see e.g
P.Ginsparg in {\it `Fields, Strings and Critical Phenomena'}
E.Brezin and J.Zinn-Justin eds., North Holland 1990. We use the notations of
this article in the present work.

\bibitem{GKO} P.Goddard, A.Kent and D.Olive {\sl Comm. Math. Phys.}
{\sl 103} (1986), 105.

\bibitem{A22} S.K Yang and H.B Zheng, {\sl Nucl.Phys.} {\bf B285[FS19]},
(1987), 410; E.B Kiritsis, {\sl J.Phys.A} {\bf 21} (1988), 297;
M.A Bershadsky, V.G Knizhnik and M.G Teitelman {\sl Phys.Lett}
{\bf 151B}, 31 (1985); G.M Sotkov and M.S Stanishkov,
{\sl Phys.Lett.} {\bf 177B},361 (1986).

\bibitem{foot1} Note that
the orbifold theory at $R=1$ is a special point
of the
Ashkin-Teller model corresponding to two decoupled Ising models, giving rise
to two Ising spin operators, of dimension $({1\over16})$.
As the Ashkin-Teller coupling between the
two Ising models is turned on, $R$ changes continuously but the spin dimension
($\eta$ exponent)
associated with these operators remain unaffected.

\bibitem{AAcoming} A.Georges and A.Sengupta, in preparation.

\bibitem{ScSc} K.Schotte and U.Schotte {\sl Phys. Rev.} {\bf 182}, 479
(1969)

\end{references}
\end{document}